# Theorizing the Socio-Cultural Dynamics of Consumer Decision-Making for Participation in Community-Supported Agriculture

Sota Takagi[1*], Yusuke Numazawa[2], Kentaro Katsube[2], Wataru Omukai[2], Miki Saijo[1], Takumi Ohashi[1,3]

[1]Tokyo Institute of Technology, 2-12-1 Ookayama Meguro-ku, Tokyo 152-8550, Japan

* takagi.s.ak@m.titech.ac.jp

[2]Eco-Pork Co. Ltd., 3-21-7 Kandanishikicho Chiyoda-ku, Tokyo 101-0054, Japan

[3]Chulalongkorn University, Phayathai Road, Pathumwan, Bangkok 10330, Thailand

**Abstract**

In the context of the urgent need to establish sustainable food systems, Community Supported Agriculture (CSA), in which consumers share risks with producers, has gained increasing attention. Understanding the factors that influence consumer participation in CSA is crucial, yet the complete picture and interrelations of these factors remain unclear in existing studies. This research adopts a scoping review and the KJ method to elucidate the factors influencing consumer participation in CSA and to theorize the consumer participation. In particular, we focus on the dynamics of individual decision-making for participation, under the premise that individuals are embedded in socio-cultural environments. We examine the decision-making process based on the seesaw of expected gains and losses from participation, along with the reflexivity to the individual and the process of updating decision-making post-participation. Our study highlights how individual decision-making for participation is influenced by relationships with others within the embedded socio-cultural environment, as well as by attachment and connection to the community. It also shows that discrepancies between expectations and experiences post-participation, and the transformation of the social capital, promote the updating of decision-making processes. In addition, among the factors identified in this study for participation in CSA, the decision to participate was heavily influenced by expectations of "variety of ingredients," suggesting that other factors such as "food education and learning opportunities," "contribution to environmental and social issues," and "connections with people and nature" had little impact. Although there are limitations, the insights gained from this study offer profound implications for stakeholders and provide valuable insights for more sustainable and efficient CSA practices.

**Keywords**

Community-supported agriculture; KJ method; scoping review; consumer participation, social capital

## 1. Introduction

There is growing interest in creating sustainable food systems as a response to environmental issues associated with the industrialization and globalization of food markets. This recent push for sustainable food systems has highlighted the importance of developing short food supply chains (SFSCs). It emphasizes strengthening the direct connections between producers and consumers and promoting the consumption of locally produced food (Princen 1997; Galli and Brunori 2013; Clapp 2015; Weber, Wiek, and Lang 2020). Community Supported Agriculture (CSA) has emerged as an important development in this movement (Woods, Ernst, and Tropp 2017; Vasquez et al. 2017). CSA represents a paradigm shift in the traditional farm-to-consumer sales model, in which consumers, often referred to as "members" or "shareholders," commit to supporting a farm operation by purchasing a share of the anticipated harvest in advance.

This innovative arrangement provides farmers with much-needed upfront capital and a guaranteed market for their produce, while consumers benefit from receiving fresh, locally produced food and an opportunity to become directly involved in the production process. CSA presents a practical solution to the economic hurdles encountered by small-scale farmers by ensuring financial stability through prepaid memberships. This stability empowers farmers to focus more on adopting and enhancing sustainable farming techniques (Worden 2004).

CSA distinguishes itself from other SFSCs such as farmers' markets not only through transactions involving the purchase of locally produced food but also by facilitating the exchange of intangible values such as interaction between producers and consumers, and opportunities to learn about agriculture, food, and the production and supply processes (Blättel-Mink et al. 2017; Egli, Rüschhoff, and Priess 2023). In addition, consumer involvement in CSA goes beyond basic operational tasks such as volunteer farming or packing. Through appropriate engagement, consumers can participate in farm management and help reduce the burden on farmers. Thus, the value offered by CSA is diverse compared to other SFSCs, and understanding which aspects of this value attract consumer participation is crucial for the sustainable operation of CSA.

Many studies revealed that consumers participate in CSA because of environmental concerns, support for local farmers, access to quality food, support for the local economy, desire to eat seasonally, and access to information about the harvest (Pole and Gray 2013; Brehm and Eisenhauer 2008; Vassalos, Gao, and Zhang 2017; Vasquez et al. 2017; Kondo 2021; Chen et al. 2019; Cone and Myhre 2000; Cox et al. 2008; Farmer et al. 2014). A choice experiment with consumers revealed that consumers who participate in farmers' markets are more positive about participating in CSA and are willing to share the risk of prepayment with farmers (Pisarn, Kim, and Yang 2020). On the other hand, these studies are often focused on specific regions and contexts and may not adequately capture the complex decision making of diverse consumers(Savarese, Chamberlain, and Graffigna 2020; Vasquez et al. 2017).

By clarifying the overall picture of these factors, we can not only provide insights that serve as a basis for formulating expansion strategies, but also make it easier to adapt the model to the characteristics of case and plan appropriate interventions. A variety of socioeconomic, psychological, and geographic attributes are thought to influence consumers' motivations and backgrounds for participating in CSA. A comprehensive understanding and systematic organization of these factors is expected to provide new insights that will contribute to the spread and development of CSA.

In this study, we aim to identify the factors that influence consumer participation in CSA by conducting a scoping review to comprehensively analyze previous research. We intend to illustrate and describe the relationships among these factors through the KJ method and theorize the consumer participation in CSA.

## 2. Materials and Methods

In this research, we undertook a scoping review to pinpoint the elements that sway consumer engagement in CSA and to understand how these elements interact. Utilizing the KJ method, factors were extracted from the reviewed literature using open coding and a theory of the consumer participation was developed by repeatedly illustrating and describing the relationships among the factors.

### 2.1. Data collection: Scoping review

For the scoping review, Web of Science All Databases, including the Web of Science Core Collection, BIOSIS Citation Index, Current Contents Connect, Data Citation Index, Derwent Innovations Index, KCI - Korean Journal Database, MEDLINE, Russian Science Citation Index, SciELO Citation Index, and Zoological Record, were utilized. This review was conducted on June 9, 2023, following the PRISMA-ScR guidelines (Tricco et al. 2018). The search query used was TS = ("Community Supported Agriculture"). To ensure the rigor and validity of the scoping review, specific inclusion and exclusion criteria were established for selecting the papers to be analyzed. The inclusion and exclusion criteria used are shown in Table 1.

**Table 1 Inclusion and exclusion criteria for scoping review.**

| Inclusion criteria | Exclusion criteria |
|---|---|
| Full-text articles | Public report, only abstract |
| Articles published in the English language | Articles published in languages other than English |
| Studies discussing consumer participation factors or motivations | Studies with consumer interventions, such as cost offset CSA |

These inclusion and exclusion criteria were adhered to, and screening was conducted by two coders: the lead author and the last author. Subsequently, the lead author thoroughly read the title, abstract and full text of each paper and conducted a full paper screening to ensure that the selected papers were relevant and contributed to the research objective of understanding the factors influencing consumer participation in CSA. In addition, to provide a more comprehensive scoping review, the papers cited by the selected articles were also screened. The same criteria shown in Table 1 were used for the screening inclusion and exclusion criteria during this process.

### 2.2. Data analysis: KJ method

In this study, after selecting papers through screening, open coding was used to extract factors from each paper that influenced consumers' participation, continuation, and withdrawal from CSA. These factors were then visualized in collaboration with co-authors. These procedures were based on the KJ method. The KJ method, which is designed to efficiently organize fragmentary information and ideas, is a tool for formulating and analyzing qualitative data. It also serves as a method for identifying essential problems and generating problem-solving ideas (Kawakita 1986; Scupin 1997). In this study, Miro, an online whiteboard application, was used for diagramming.

The key feature of the KJ method is that it enables users to decontextualize text from the target data and then recontextualize it through a process of grouping and articulating sentences combined with visualization. Some case studies were included in the scoping review conducted for this study, it is considered that specific socio-economic and geographical factors of the countries may influence the motivations and backgrounds of consumers participating in CSA. Therefore, through this decontextualization and recontextualization approach, it is possible to systematically organize individual pieces of information, clarify information on a broader scale, and discover new perspectives and relationships to construct inferences. The constructivist design of this method facilitates a deep understanding of complex information and promotes the construction of inferences that are firmly rooted in the context of the research area. The ability of the KJ method to provide practical and specific inferences useful in such practices is what made it suitable for this study.

The theorization of the consumer participation through illustrating and description in this study comprises four major steps as shown in Fig. 1: Step 1. Label making, Step 2. Label grouping, Step 3. Chart making, Step 4. Written explanation. In our study, the Step 1 and Step 4 processes were conducted by the lead author, while Steps 2 and 3 were collaboratively executed by the lead author and the last author. The iteration of Steps 2 to 4 was informed by validation and feedback from other co-authors, and the chart was modified as needed.

This process of illustrating and revising possesses characteristics akin to peer debriefing techniques (Janesick 2015). Peer debriefing is employed as a method to enhance the reliability and credibility of research, allowing for the fortification of theoretical robustness through collaborative deliberation and revision among researchers. In our study, this process ensured that the theory aptly reflected the data. Discrepancies in opinions were resolved through consensus. Consequently, the refined chart and explanation were adopted as the central theory of our research.

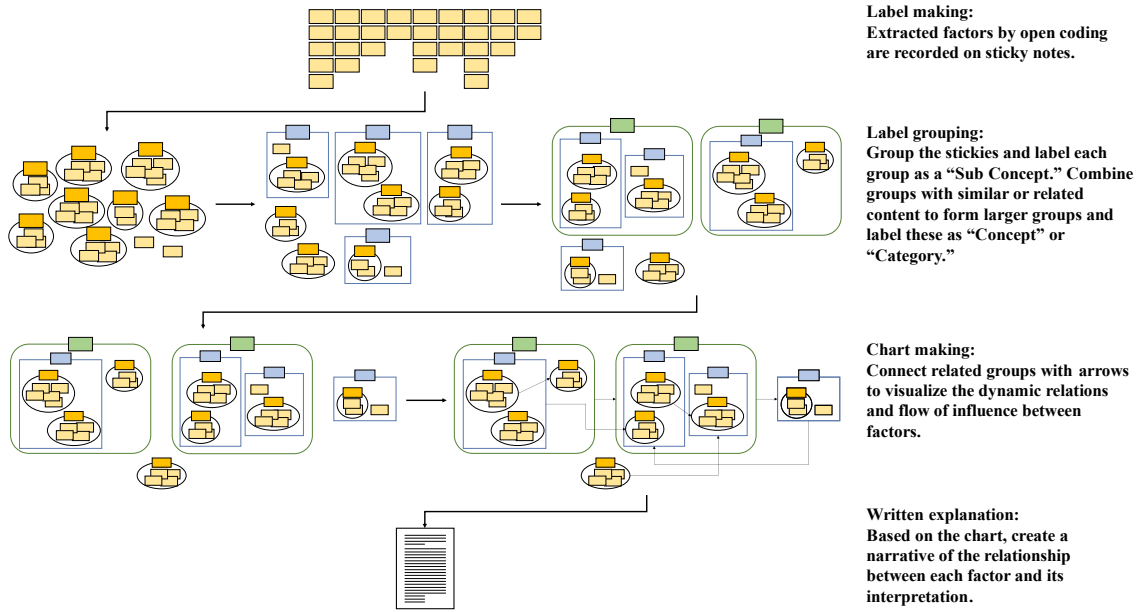

**Label making:**
Extracted factors by open coding are recorded on sticky notes.

**Label grouping:**
Group the stickies and label each group as a "Sub Concept." Combine groups with similar or related content to form larger groups and label these as "Concept" or "Category."

**Chart making:**
Connect related groups with arrows to visualize the dynamic relations and flow of influence between factors.

**Written explanation:**
Based on the chart, create a narrative of the relationship between each factor and its interpretation.

**Fig. 1 Theorizing Process using the KJ method.**

### 2.3. Researcher Characteristics and Reflexivity

The research team comprised six researchers with diverse backgrounds and expertise. The lead author, the principal investigator, has experience visiting small farms, volunteering, and interacting with producers. This experience, considering the nature of this study's focus on consumer participation in CSA, is noteworthy. In addition, the authors' thorough reading of the full texts of the articles selected through screening during the research process influenced their understanding of the context of CSA research. These characteristics may have influenced our approach to open coding and grouping in this study, potentially providing unique perspectives in the analysis of factors affecting consumer participation. The second author, who serves as a PR/marketing representative for a pork marketing company with a focus on sustainability, brings a unique perspective to this research. His extensive experience in reporting on producers' efforts toward sustainable production and regional contributions, and in disseminating information to consumers through websites and the media, has provided practical insights and had a profound impact on the data analysis and theoretical model of this study. The third author, as CMO of the same company, has experience in consumer-facing business, launching e-commerce sites, and product development, providing a broad understanding of both producers and consumers. In addition, his experience in initiating community spaces as a planner provides a comprehensive understanding of community formation and the participation process, which significantly influences the theoretical construction. The fourth author, actively involved in marketing, brings extensive experience in advocating social contributions through food to consumers, providing a broad understanding of consumer perspectives. In addition, his experience as a consumer using a farm-to-door vegetable service for five years provides a unique perspective on the consumer participation in CSA, which greatly influences the theory construction. The fifth author, while not possessing extensive knowledge in agricultural industry, holds a Ph.D. in Applied Linguistics. This unique knowledge considerably influenced the process of data collection and interpretation, especially the process of narrativization. The last author of this study possesses extensive experience in conducting scoping reviews across various fields, designing and managing workshops utilizing the KJ method, and a wealth of knowledge in data analysis using the KJ method. In addition, the last author's research on smart livestock technology, grounded in the iterative and reflexive approaches of human-centered design, significantly influenced the theorization of the consumer participation in this study.

Following the constructivist paradigm, our diverse backgrounds and experiences influenced each stage of the research process, from factor extraction to theorizing the consumer participation through recontextualization. Each researcher brought potential biases to the study. To mitigate these biases, we shared information among co-authors and clarified different perspectives on the data to achieve a balanced interpretation. In this process, the use of the KJ method for visualization and storytelling played a critical role in facilitating co-author understanding. Collaboration and discussion among co-authors with diverse backgrounds helped to prevent unique perspectives and biases from influencing the results and provided new insights for a deeper understanding of consumer participation in CSA.

## 3. Results

### 3.1. Literature selection process and target papers: A Scoping review

Fig. 2. shows the flowchart of the scoping review process according to the PRISMA-ScR guidelines. We initially identified 505 papers from an extensive search across the Web of Science All Databases. These papers underwent a detailed screening process based on specific inclusion and exclusion criteria, focusing on original research articles published in English-language open access journals. This screening narrowed the selection to 183 papers.

The lead and last authors then reviewed the titles and abstracts of these 183 papers, selecting those that aligned with the criteria detailed in Table 1. This process further refined the selection to 46 papers. Of the 46 papers, 6 were excluded, including 1 "Data in Brief," 1 irrelevant to CSA, and 4 unrelated to consumer participation factors.

Further, we extended our screening to include 69 unique papers cited in the 40 papers. This secondary screening, following the same criteria, led to the addition of 21 more papers, culminating in a total of 61 papers included in our study. This comprehensive process ensured a thorough and relevant collection of research for our analysis.

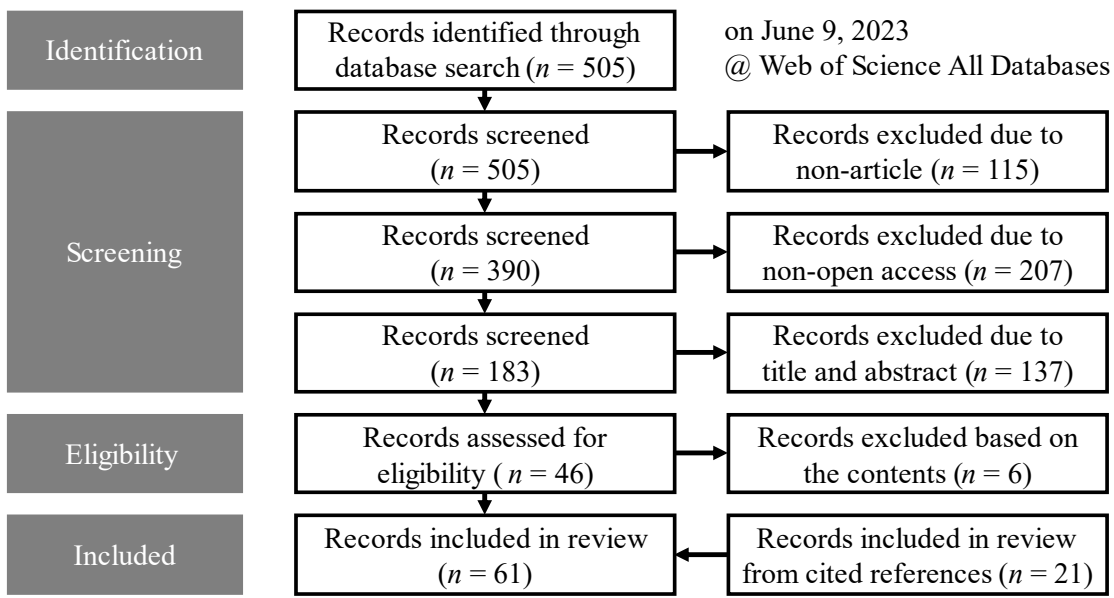

**Fig. 2 PRISMA-ScR Flowchart.**

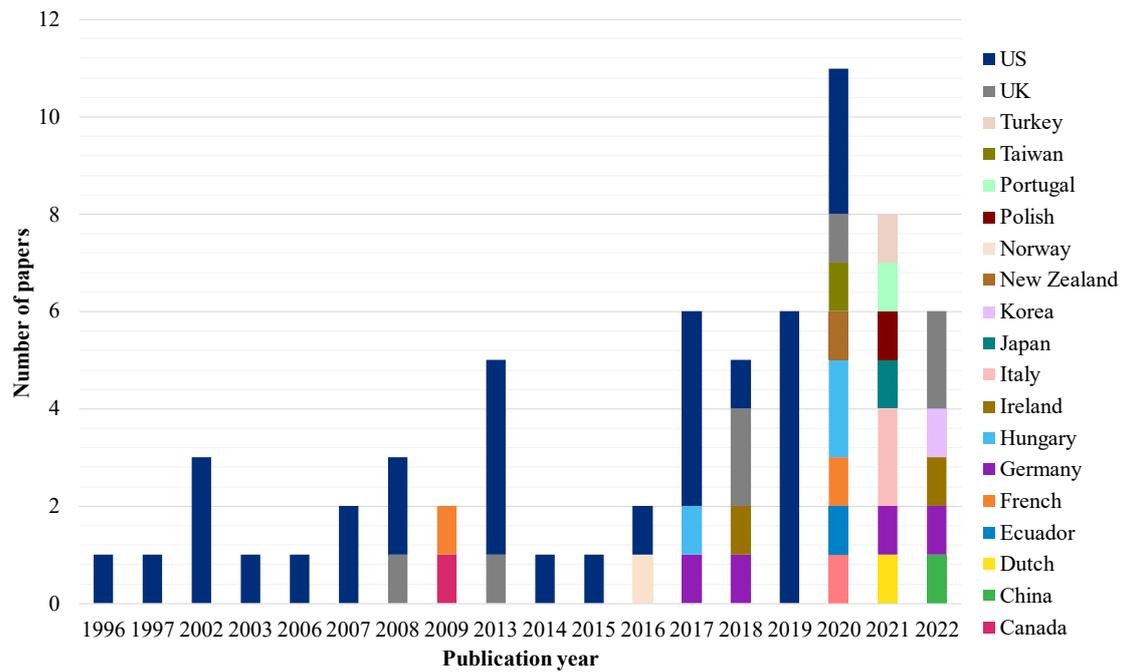

**Fig. 3 Publication years and countries studied over the years.**

Fig. 3 shows the year of publication and country of coverage for the 61 papers analyzed in the review. The total number of papers in the graph is 65, as some papers cover more than one country. The number of papers on factors contributing to consumer participation in CSA shows an increasing trend: of the 40 papers published before 2019, 29 were studies in the US, while the increase in studies in other countries after 2020 indicates that the entry of non-US researchers into the field is responsible for the increase in the number of papers. It also shows the increased interest in vulnerabilities and risks of the current food supply chain brought about by COVID-19, with 2 out of 11 papers in 2020, 3 out of 6 in 2021, and 4 out of 5 in 2022 mentioning COVID-19. This also suggests that the pandemic has increased interest in CSA and changed added a new focus for research.

### 3.2. Factors influencing consumer participation in CSA

To theorize the consumer participation, we used the KJ method to illustrate and describe the data. First, we applied open coding to the 61 papers selected through screening, extracting 306 factors. These factors were recorded on sticky notes for further illustration and narrative development. This process identified 6 categories, 23 concepts, and 68 sub concepts.

We then created a core category that summarized the extracted factors, categories, concepts, and sub concepts. A correspondence table detailing all extracted factors along with their relationship to the core category, categories, concepts, and sub concepts, including source information, is provided in the Appendix. In abstracting the factors that influence consumer participation, it became clear that they largely fall into two categories: "Socio-Cultural Environment" and "Seesaw of Gain and Loss."

The "Socio-Cultural Environment" illustrates that individuals are embedded in a socio-cultural context, such as networks of relationships with others, that shapes their attitudes and behaviors. These attitudes and behaviors are influenced at multiple levels, including the family/peer level, the local environment and community, and national agricultural framework conditions. The concept of the "Seesaw of Gain and Loss" refers to decision-making influenced by the balance between expected gains and losses from participation in CSA. This includes not only individual perceptions of gains and losses but also the acceptance of risks associated with CSA, such as financial and time commitments, which are also influenced by the socio-cultural environment. The specifics of each of these components are discussed in the following sections.

### 3.2.1. Socio-Cultural Environment

Tables 2 through 5 present the composition and explanation of the socio-cultural environment at each level. We categorized this into individual, family/peer, local environment and community, and national agricultural frameworks, and defined core categories for each of these components.

**Table 2 Socio-Cultural Environment: Individual.**

| Core Category | Description | References |
|---|---|---|
| Knowledge and Experience | Knowledge and experience with CSA or similar direct marketing. | (Pisarn, Kim, and Yang 2020; Vassalos, Gao, and Zhang 2017; Yu et al. 2019; Hanson et al. 2019; Morgan et al. 2018; Diekmann, Gray, and Thai 2020) |
| Skills | The level of experience in cooking fresh food, the cooking environment, and the ability to handle unfamiliar ingredients. | (Hanson et al. 2019; O'Neill et al. 2022; Andreatta, Rhyne, and Dery 2008; Sitaker et al. 2020; Zepeda and Li 2006; Galt, Bradley, et al. 2019; Lee 2022; Rossi et al. 2017) |
| Attitude towards Food | Attitude towards food, including particularities in obtaining ingredients, and the effort and time dedicated to food preparation. | (Cotter et al. 2017; Vasquez et al. 2016; Galt, Van Soelen Kim, et al. 2019; April-Lalonde et al. 2020; Schmutz et al. 2018; Pole and Kumar 2015; Schnell 2013; Farmer et al. 2014; Ostrom 2007; O'Hara and Stagl 2002; Farnsworth et al. 1996; Opitz et al. 2017; Kolodinsky and Pelch 1997; Hvitsand 2016; Durrenberger 2002; Chen et al. 2019; Cox et al. 2008; Yu et al. 2019; Diekmann, Gray, and Thai 2020; Bernard, Bonein, and Bougherara 2020; Birtalan, Bartha, et al. 2020; Brehm and Eisenhauer 2008; Vassalos, Gao, and Zhang 2017; Sitaker et al. 2020; Birtalan, Neulinger, et al. 2020; Galt, Bradley, et al. 2019; Wang et al. 2021; Morgan et al. 2018; Zoll et al. 2018; Gorman 2018; Perez, Allen, and Brown 2003; Bakos 2017; Zepeda and Li 2006) |
| Attitude towards Environment | Attitude towards environmental issues and social issues at the national or regional level. | (Vassalos, Gao, and Zhang 2017; Andreatta, Rhyne, and Dery 2008; Rossi et al. 2017; Kolodinsky and Pelch 1997; Schmutz et al. 2018; Schnell 2013; Ostrom 2007; Hvitsand 2016; Durrenberger 2002; Chen et al. 2019; Cox et al. 2008; Birtalan, Bartha, et al. 2020; Zoll et al. 2018; Bougherara, Grolleau, and Mzoughi 2009) |
| Attitude towards Health | Attitude towards healthy lifestyles and activities, including diet and exercise. | (Andreatta, Rhyne, and Dery 2008; Cox et al. 2008; Vassalos, Gao, and Zhang 2017; O'Hara and Stagl 2002; Birtalan, Bartha, et al. 2020; April-Lalonde et al. 2020; Galt et al. 2017) |

**Table 3 Socio-Cultural Environment: Family/Peer.**

| Core Category | Description | References |
|---|---|---|
| Family/Peer Environment and Value | Family/peer situations that require attention to health, or a shared household understanding of the need for health awareness and effort in food preparation. | (Zepeda and Li 2006; April-Lalonde et al. 2020; Kolodinsky and Pelch 1997; Birtalan, Neulinger, et al. 2020; Galt et al. 2017) |

**Table 4 Socio-Cultural Environment: Local Environment and Community.**

| Core Category | Description | References |
|---|---|---|
| Connection and Attachment to the Community | The presence and degree of connection or sentiment towards the community. | (Vassalos, Gao, and Zhang 2017; Sitaker et al. 2020; Pole and Kumar 2015; Farmer et al. 2014; Ostrom 2007; Opitz et al. 2017; Kolodinsky and Pelch 1997; Hvitsand 2016; Chen et al. 2019; Brehm and Eisenhauer 2008; Gorman 2018; Perez, Allen, and Brown 2003; Bakos 2017; Galt et al. 2017; Kondo 2021; Pole and Gray 2013; Kato 2013) |
| Culture | The presence of traditional landscapes, farming practices, or crops in the region. | (Opitz et al. 2017; Gorman 2018; Schnell 2013) |
| Local Norms | The presence and degree of norms in the community regarding the support of farmers and the community. | (Morgan et al. 2018; Diekmann, Gray, and Thai 2020; Schnell 2013; O'Hara and Stagl 2002; Ostrom 2007; Farnsworth et al. 1996; Hvitsand 2016; Durrenberger 2002; Brehm and Eisenhauer 2008; Wang et al. 2021; Zoll et al. 2018; Perez, Allen, and Brown 2003; Galt et al. 2017; Thompson and Coskuner-Balli 2007; Zoll, Specht, and Siebert 2021; Ravenscroft et al. 2013; Sharp, Imerman, and Peters 2002) |

**Table 5 Socio-Cultural Environment: National Agricultural Framework Conditions.**

| Core Category | Description | References |
|---|---|---|
| Trends and Maturity of CSA | Popularity and maturity of CSA activities. | (Ostrom 2007; Farnsworth et al. 1996; Hvitsand 2016; Kondo 2021; Feagan and Henderson 2009; Bonfert 2022; Pelin and Murat 2021) |
| Support Systems and Organizations | The existence of efforts and organizations that provide subsidies and grants to promote CSA at national and regional levels. | (Yu et al. 2019; Savarese, Chamberlain, and Graffigna 2020) |
| Agriculture Policy | Policy factors influencing CSA implementation. | (Mert-Cakal and Miele 2020; Plank, Hafner, and Stotten 2020; Durrenberger 2002; Yu et al. 2019; Galt, Van Soelen Kim, et al. 2019) |

### 3.2.2. Seesaw of Gain and Loss

Tables 6 and 7 detail the composition and explanation of the seesaw of gain and loss, focusing on expected gain and expected loss. Four types of gains and three types of losses were identified and extracted as core categories.

**Table 6 Seesaw of Gain and Loss: Gain.**

| Core Category | Description | References |
|---|---|---|
| Food Education and Learning Opportunities | The presence and variety of learning opportunities regarding food ingredients and agriculture provided by producers. | (Morgan et al. 2018; Andreatta, Rhyne, and Dery 2008; April-Lalonde et al. 2020; Opitz et al. 2017; Kolodinsky and Pelch 1997; Hvitsand 2016; Zoll et al. 2018; Feagan and Henderson 2009; Savarese, Chamberlain, and Graffigna 2020; Zepeda, Reznickova, and Russell 2013) |
| Contribution to Environmental and Social Issues | Awareness regarding solving environmental issues and social problems at the national and regional levels through participation in CSA. | (Vassalos, Gao, and Zhang 2017; Ostrom 2007; Farnsworth et al. 1996; Cox et al. 2008; Zoll et al. 2018; Zoll, Specht, and Siebert 2021; Savarese, Chamberlain, and Graffigna 2020) |
| Connections with People and Nature | Meeting people through CSA participation and interacting with nature and animals through work on the farm. | (Galt, Bradley, et al. 2019; Pole and Kumar 2015; Schnell 2013; O'Hara and Stagl 2002; Hvitsand 2016; Cox et al. 2008; Brehm and Eisenhauer 2008; Wang et al. 2021; Zoll et al. 2018; Gorman 2018; Pole and Gray 2013; Bonfert 2022; Mert-Cakal and Miele 2020; Zepeda, Reznickova, and Russell 2013; Furness et al. 2022; Piccoli, Rossi, and Genova 2021) |
| Variety of Ingredients | The content of the share, including the amount and type of food ingredients, and the frequency of sharing. | (Pisarn, Kim, and Yang 2020; Yu et al. 2019; Hanson et al. 2019; Morgan et al. 2018; O'Neill et al. 2022; Andreatta, Rhyne, and Dery 2008; Sitaker et al. 2020; Galt, Bradley, et al. 2019; Vasquez et al. 2016; Galt, Van Soelen Kim, et al. 2019; Pole and Kumar 2015; Schnell 2013; O'Hara and Stagl 2002; Ostrom 2007; Opitz et al. 2017; Kolodinsky and Pelch 1997; Durrenberger 2002; Bernard, Bonein, and Bougherara 2020; Wang et al. 2021; Zoll et al. 2018; Gorman 2018; Perez, Allen, and Brown 2003; Bougherara, Grolleau, and Mzoughi 2009; Galt et al. 2017; Pole and Gray 2013; Thompson and Coskuner-Balli 2007; Sharp, Imerman, and Peters 2002; Plank, Hafner, and Stotten 2020; Zepeda, Reznickova, and Russell 2013; Sitaker et al. 2019) |

**Table 7 Seesaw of Gain and Loss: Loss.**

| Core Category | Description | References |
|---|---|---|
| Complicated Relationships | Constraints and obligations within the community due to CSA participation, as well as complications in communicating with members and farmers. | (Ravenscroft et al. 2013; Poças Ribeiro et al. 2021; Kondo 2021; Medici, Canavari, and Castellini 2021) |
| Money | The costs associated with purchasing a share in the CSA and the difference in spending compared with previous food procurement costs. | (Yu et al. 2019; Andreatta, Rhyne, and Dery 2008; Sitaker et al. 2020; Cotter et al. 2017; O'Hara and Stagl 2002; Kolodinsky and Pelch 1997; Perez, Allen, and Brown 2003; Kato 2013; Plank, Hafner, and Stotten 2020; McGuirt et al. 2020) |
| Time | The additional time required for picking up the shares and preparing unfamiliar ingredients or fresh produce. | (Perez, Allen, and Brown 2003; Sitaker et al. 2020; 2019; Morgan et al. 2018; Plank, Hafner, and Stotten 2020; Galt, Bradley, et al. 2019; Bakos 2017; Kato 2013) |

### 3.3. Theory and narrative of the consumer participation

Considering the two factors of "Socio-Cultural Environment" and "Seesaw of Gain and Loss," we created a theoretical diagram, as shown in Fig. 4. Below is the narrative description of the theoretical diagram. The Core categories are indicated by [Core Category], and categories, concepts, and sub concepts are <u>underlined</u>.

Consumer participation in CSA is significantly influenced by individual [Attitude towards Food], such as dedication to ingredient quality and the effort put into cooking. Additionally, [Attitude towards Health], formed through maintaining a healthy diet and physical activity, and [Attitude towards Environment] regarding environment and social issues at the national and regional levels also play a crucial role.

Furthermore, an individual's [Skills] in preparing fresh produce and [Knowledge and Experience] in direct marketing, including CSA, also affect participation.

However, these factors necessitate a thorough consideration of the premise that individuals are embedded in a Socio-Cultural Environment. This environment comprises not only the Individual but also the Family/Peer, Local Environment and Community, and National Agricultural Framework Conditions.

At the Family/Peer level, the presence of family/peer members who require action to improve their health and a collective understanding within the family/peer group about health, categorized under [Family/Peer Environment and Value], are believed to influence an individual's attitudes and behaviors.

The family/peer group's engagement in <u>leading a life with health considerations</u> and their <u>values towards food as a family/peer</u>, when imparted to an individual, are thought to affect their [Attitude towards Health] and [Attitude towards Food]. This is because sharing meals within the same household and upholding values like health practices through food, including the <u>ability to devote effort to food preparation</u>, are interpreted as contributing to shaping an individual's attitudes and behaviors. Additionally, families/peer groups that <u>can/do cook fresh food at home</u>, are also considered to affect the development of [Skills].

At the Local Environment and Community level, [Connection and Attachment to the Community] and [Local Norms] can shape the <u>desire to consume local ingredients</u> and <u>enthusiasm for supporting farmers and the local area</u>, which influence the individual's and family/peer group's [Attitudes towards Food] and [Attitudes towards Environment]. In addition, <u>having experience participating in CSA</u> or <u>using farmers' markets</u> and having the connections with farmers is assumed to influence the frequency of participation or use, affecting the degree of [Knowledge and Experience]. In the presence of a community environment in which <u>word of mouth spreads easily</u>, or in which the region is involved in <u>landscape preservation and environmental management</u> and emphasizes the [Culture], information about CSA and values about the community and nature are more likely to be shared with individual and family/peer group. This not only develops <u>values towards food as a family/peer</u> but also the values fostered also influence the individual.

At the National Agricultural Framework Conditions level, there are macro-level factors related to consumers' access to CSA as a precondition, such as the existence of [Agriculture Policy] and [Support Systems and Organizations] that influence the implementation of CSA, and [Trends and Maturity of CSA] that affect awareness of CSA.

In such a socio-cultural environment, individual decision-making is based on the seesaw of gain and loss. When the balance between expected gain and expected loss tips towards expected gain, consumers participate in CSA. Expected gains include [Food Education and Learning Opportunities], [Variety of Ingredients], and [Connections with People and Nature], as well as [Contribution to Environmental and Social Issues] provided through CSA participation. Expected losses include [Complicated Relationships] associated with <u>expectation of human connections through CSA participation</u> with consumers and producers. If there are <u>many restrictions within the community</u>, consumers may perceive human relationships as a loss. There is also a loss of [Money] and [Time], such as a lack of <u>feeling the fairness in the content and price of the share</u> and <u>high accessibility</u> to pick-up points.

The balance is influenced by individual perceptions and levels of risk acceptance. These levels of <u>recognition of the benefits of taking risks</u> and risk acceptance are in turn influenced by the socio-cultural environment.

Participation in CSA leads to influences from the values and norms within the CSA

community, transforming an individual's [Knowledge and Experience], [Skills], and [Attitudes]. These transformations in turn affect the socio-cultural environment and the seesaw of gain and loss. Because the transformed socio-cultural environment and seesaw of gain and loss once again influence individual decision-making, participation in CSA is a reflexive process. In addition, decisions are further influenced by the gap between expected gains and experienced losses from participation, such as <u>financial and time cost performance</u>, as well as new perceived gains and losses, and changes in risk acceptance based on these factors.

Thus, under the premise that consumers are embedded in a socio-cultural environment, they are influenced by that environment as well by the reflexive and updating processes associated with individual actions that lead to decision-making and participation in CSA.

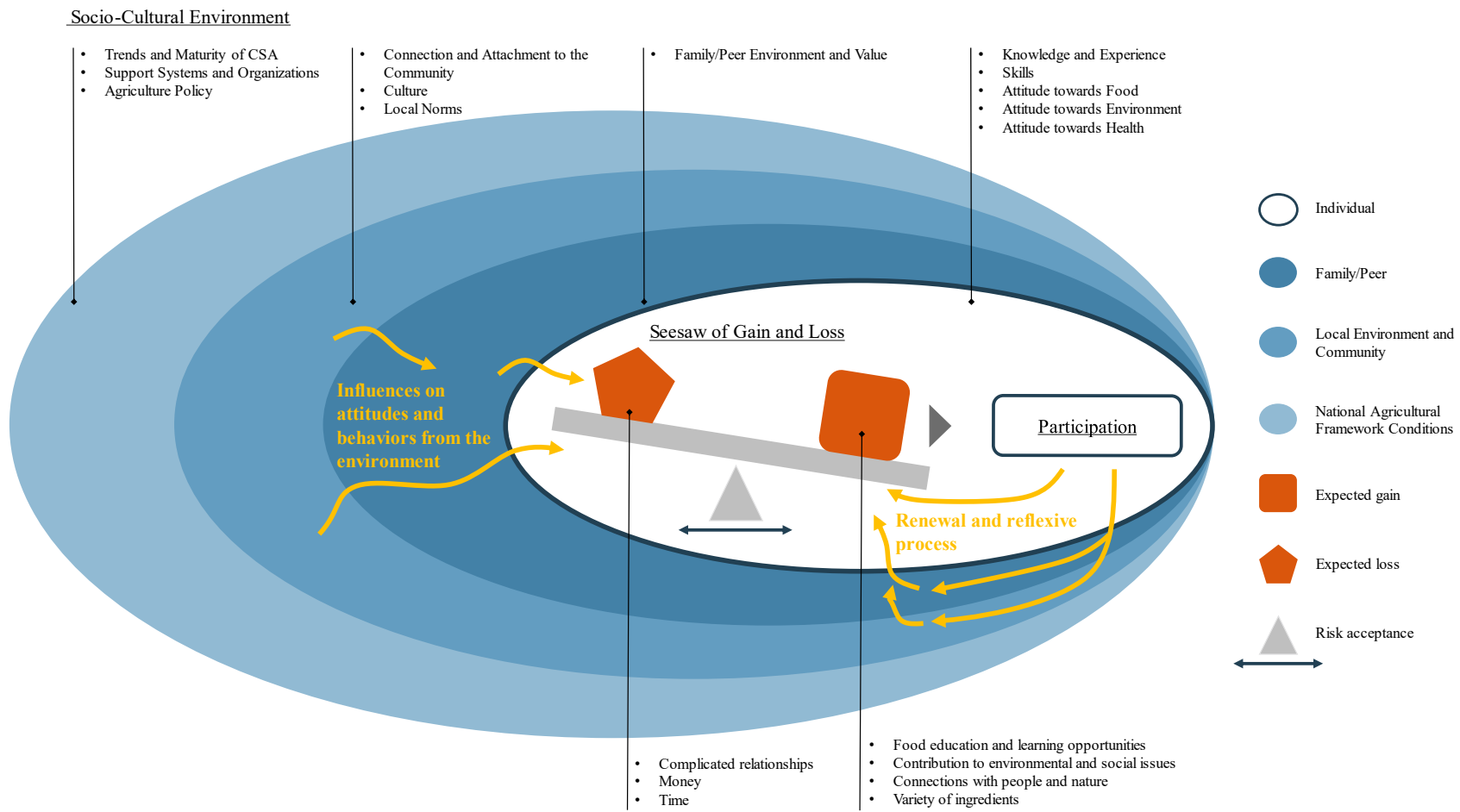

**Socio-Cultural Environment**

- Trends and Maturity of CSA
- Support Systems and Organizations
- Agriculture Policy

- Connection and Attachment to the Community
- Culture
- Local Norms

- Family/Peer Environment and Value

- Knowledge and Experience
- Skills
- Attitude towards Food
- Attitude towards Environment
- Attitude towards Health

Seesaw of Gain and Loss

Influences on attitudes and behaviors from the environment

Participation

Renewal and reflexive process

- Complicated relationships
- Money
- Time

- Food education and learning opportunities
- Contribution to environmental and social issues
- Connections with people and nature
- Variety of ingredients

- ⬭ Individual
- ⬤ Family/Peer
- ⬤ Local Environment and Community
- ⬤ National Agricultural Framework Conditions
- ■ Expected gain
- ⬠ Expected loss
- ▲ Risk acceptance
- ↔

**Fig. 4 CSA Participation Model: Socio-Cultural Dynamics and Embeddedness.**

# 4. Discussion

## 4.1. Theoretical implication

### 4.1.1. Influence of Socio-Cultural environment on the individual

Our theoretical model posits that individuals are embedded in a socio-cultural environment and that this environment is divided into four levels: National Agricultural Framework Conditions, Local Environment, Community and Family/Peer and Individual. These influence the formation of individual attitudes and behaviors. In particular, the relationship between the family/peer group and the individual is closely intertwined, and the decision-making processes and values within the family/peer group strongly influence the individual. For example, family decision-making falls into three main categories: husband-dominant, wife-dominant, and joint decision-making by the couple (Jenkins 1978; Filiatrault and Ritchie 1980; Nichols and Snepenger 1988; Fodness 1992). Thus, considering this in a family/peer group, if the dominant party adopts information and values through word-of-mouth or connections within the local environment and community and shares them at home, or bases his or her decision to participate in CSA on this information, the attitudes and behaviors of the non-dominant individual may be somewhat coerced. Conversely, if the non-dominant party is involved in CSA and this results in increased effort within the household, it is conceivable that the dominant party's decision could lead to the formation of a decision to withdraw from CSA.

This influence of the surrounding environment on the formation of an individual's attitudes and behaviors is also evident in the socio-ecological model. This model highlights how different levels of an individual's environment, from immediate family/peers to broader community and societal factors, play a crucial role in shaping their behaviors and attitudes (Dahlberg and Krug 2006). However, a point of difference between the socio-ecological model and our theory is the significant influence of community-level factors, such as the local environment and community, on the individual, in addition to direct relationships such as family/peer. This includes not only local farmers and people within the community but also the culture and natural values inherent to the region itself. Our study suggests that strong connections and attachments to such cultural and natural aspects of a region, represented as "Connection and Attachment to the Community" (Core Category), play a role in forming the motivation to support the local community. This aligns with the concept that in "locations with natural surroundings," both place attachment and an individual's Environmental Identity (Clayton 2003) contribute to place-specific pro-environmental behaviors. This connection indicates that an individual's sense of belonging and identity related to the environment play a crucial role in fostering behaviors that are beneficial to that particular natural setting (Naiman, Allred, and Stedman 2021; Tonge et al. 2015; Halpenny 2010).

As stated above, individuals are embedded in their socio-cultural environment, and within this environment, it is said that through connections with others, individuals sometimes rely on the trust and intuitive judgments within their networks than on rationality (M. Granovetter 1985). Therefore, it can be assumed that when individuals establish strong relationships with their family/peer group or residents and farmers in the local environment and community, or when they have a strong attachment to their region, attitudes and behaviors that prioritize the benefits of their family/peer group or local environment and community are likely to be formed.

In conclusion, participation in CSA is not an isolated act by an individual but is deeply influenced by a broader context, including the individual's network with others such as family/peer group and people in the local community, as well as the physical environment in which they live. However, there are limitations in explaining family/peer and community interrelationships in this study. In addition, there is a paucity of previous research that focuses on family/peer dynamics and the impact of the physical environment, such as nature, on participation. Therefore, to gain a deeper understanding of how these relationships influence pro-environmental behaviors and contributions to the community, and ultimately lead to participation in CSA, further analysis is needed in future research.

### 4.1.2. Balancing Expected Gains and Losses: Consumer Decision-Making Dynamics

Our theory shows that individuals' decisions are based on the balance between expected gains and expected losses for consumers. As explained in the previous section, individuals form attitudes and behaviors under the influence of their socio-cultural environment. These attitudes and behaviors influence the balance between expected gain and expected loss and risk acceptance in individual decision making.

We identified sub concepts such as "feeling the fairness in the content and price of the share" (sub concept), representing monetary gains and losses. In addition, time-related losses such as "increasing time/effort spent on cooking vegetables at home" (extracted factor) and "pickup points in inconvenient locations" (extracted factor) were also mentioned as reasons for withdrawal (Galt, Bradley, et al. 2019; Birtalan, Neulinger, et al. 2020). Based on these findings, the following two factors influence consumer decision-making: discrepancies between expected gains and experienced gains, and discrepancies between expected losses and experienced losses.

Previous studies have shown that having many unwanted products, or few types of share contents available, is one reason members leave CSA programs (Flora and Bregendahl 2012). Moreover, a choice experiment conducted with consumers in Connecticut interested in joining a CSA revealed that regardless of prior CSA experience, there is a higher willingness to pay for shares that offer financial compensation when the harvest is poor (Yu et al. 2019). These findings and the result of our study on monetary gains and losses align with prior study that consumers who participate in CSA tend to be risk-averse and avoid inequality (Bernard, Bonein, and Bougherara 2020).

Based on the perspective of Expectation Disconfirmation Theory (Oliver 1980), receiving a smaller share than expected or requiring more effort reduces consumer satisfaction, the "variety of ingredients" and "money" (Core category) are considered key factors that directly link to consumer satisfaction with participation.

In addition to providing access to ingredients, CSA also highlights offering nonmaterial gains that consumers expect from participating in CSA: "food education and learning opportunities," "connections with people and nature," and "contribution to environmental and social issues" (Core category). However, previous studies have shown that among consumers' motivations for participation, interest in community building and learning about agriculture ranked lower than access to ingredients (Ostrom 2007). Furthermore, while some consumers clearly recognize that participating in a CSA contributes to environmental sustainability, they have discontinued their memberships due to the effort required to cook the vegetables, which is consistent with the loss referred to as "time" (Core category) identified in our study. From these findings, it is suggested that the decision making to participate in CSA is greatly influenced by the expectation of access to "variety of ingredients" (Core category), and the impact of other categories such as "food education and learning opportunities," "contribution to environmental and social issues" and "connections with people and nature" (Core category) is generally small. Therefore, it is considered crucial to reduce the gap between financial expectations and experiences for the long-term commitment of consumers, for example, some farmers address this gap by offering a la carte options and compensation during poor harvests (Flora and Bregendahl 2012).

However, according to the Expectation Disconfirmation Theory, because positive discrepancies lead to satisfaction, it is conceivable that consumers who initially had little interest in building a community and perceived "complicated relationships" (Core category) as a loss associated with participation could develop positive relationships and build a community through interactions with people and nature. This could create a positive gap between expectation and experience, leading to a gain referred to as "connections with people and nature" (Core category). This can lead to an updating of the balance of the seesaw of gain and loss and risk acceptance, potentially leading to greater commitment and settling into the community. Previous research has shown that the effect of the perceived value of a service is fully mediated by satisfaction, leading to repurchase intentions (Patterson and Spreng 1997). Therefore, by minimizing the monetary gap and providing intangible values that lead to satisfaction, it is conceivable that this could result in long-term commitment.

In conclusion, our theory underscores that consumer decision-making in CSA is fundamentally influenced by the interplay of expected gains and losses, which are deeply rooted in socio-cultural factors. The results of this study and previous research suggest that among the participation factors, the expectation of access to "variety of ingredients" (Core category) has a significant impact compared to other core categories in terms of "gain." On the other hand, some enthusiastic members are attracted to CSA not only for tangible gains such as fresh produce but also for nonmaterial gains such as educational opportunities, community connections, and the ability to contribute to environmental and social problems. Critical to this decision-making process is the gap of consumer expectation and experience with the values and offerings of CSA producers. This alignment or discrepancy between expected and experienced gains or losses, viewed through the lens of Expectation Disconfirmation Theory, updates the seesaw and plays a critical role in consumer satisfaction.

The study acknowledges significant limitations, primarily its focus exclusively on the consumer perspective, which overlooks crucial insights from producers. Understanding the producer's view of consumer expectations and the producer's ability to meet those expectations is essential to further refining the model in this study and requires further research.

### 4.1.3. Interplay of Participation and Social Capital in CSA

Our study suggests the development and transformation of social capital associated with consumer participation. When consumers participate in farm work, volunteer activities, and farm management in a CSA, there is the possibility of creating a high-density network among consumers, producers, and other stakeholders. This can lead to the formation of social capital (Coleman 1988). There are three forms of social capital are examined: obligations and expectations, information channels, and social norms. Social capital consisting of such a strong network is classified as bonding social capital (Patulny and Svendsen 2007). Because these factors increase consumer commitment, bonding social capital is important from the perspective of sustainable CSA. However, the presence of "overly strong relationships within the community" (extracted factor), "strict constraints within the community" (extracted factor), and "a lot of work on the farm" (extracted factor) suggest that strong relationships, obligations and social norms within the CSA community can also encourage consumers to leave and inhibits CSA development (Poças Ribeiro et al. 2021; Ravenscroft et al. 2013).

Previous research has shown that members who experience more social capital benefits through the participation in the CSA community are more likely to remain members (Flora and Bregendahl 2012). Furthermore, enthusiastic participants significantly change their dietary habits(Feagan and Henderson 2009), and in California, a study found that 82% of households whose diets had changed as a result of participating in CSA expressed a desire to continue the membership (Perez, Allen, and Brown 2003). This suggests that members with higher levels of engagement improve their diets and participate in a cyclical process that encourages continued engagement. From our study and previous research, it can be inferred that bonding social capital plays a crucial role in people's commitment and that the development and transformation of bonding social capital within the community can impact both long-term participation and withdrawal.

In addition, prior studies suggest that bridging social capital arises from weak networks formed (M. S. Granovetter 1973; Burt 1992) and has the potential to spread change from immediate communities to peripheral communities (Furness et al. 2022). The operation of CSA requires consumer engagement and linkages with external organizations when there is a lack of resources in the community, such as volunteer labor or knowledge, or when solving problems to achieve future goals, such as increasing scale. The importance of investing in bridging social capital is also mentioned because it facilitates outreach to consumers with different resources outside the community and the incorporation of external funding (Pelin and Murat 2021).

Previous research has revealed that there is CSA operating at full capacity with a waiting list for new members, despite having low levels of both bonding social capital and bridging social capital. However, this CSA primarily communicate via email with very little face-to-face interaction, and their purpose is limited to providing local households with organic vegetables (Furness et al. 2022). This suggests that for CSA aiming to continue operations by continually acquiring new members, much like a typical transactional relationship between producers and consumers, the construction and transformation of social capital are not deemed important. Nevertheless, as evidenced by the results of our study, CSA offer diverse values beyond mere food transactions, such as opportunities for interaction with people and food education, which also serve as factors for consumer participation. Therefore, the sustainable operation of CSA that offer diverse values requires the building of trust and long-term commitment from consumers, indicating the importance of both dense relationships within the community that build bridging social capital and weaker connections with external consumers and stakeholders that establish bridging social capital.

In conclusion, the sustainability and evolution of CSA are significantly influenced by social capital. Although numerous studies have examined the creation and importance of social capital in CSA, the changes in social capital resulting from individual participation and its reflexive nature remain largely unexplored. Furthermore, the synergies between bonding social capital and bridging social capital are not clear. Moreover, the scope of this study did not extend to identifying specific changes in decision-making processes attributable to the transformation of social capital. Consequently, future research should delve into the roles, actions, and evolution of individuals within communities, using methodologies such as action research and in-depth interviews.

## 4.2. Practical implication
### 4.2.1. Insight for CSA promotors and farmers willing to practice CSA

The results of our study underscore that individuals are embedded in a socio-cultural environment in which decision-making within the family/peer group and relationships with the community are critical factors in shaping attitudes and behaviors. CSA promoters and farmers interested in practicing CSA need to consider decision-making based on the lifestyles and values of the entire family/peer group and increase points of contact with them. Extracted factors such as "CSA having collaborations with other institutions (universities, schools, social movements)" (extracted factor) and "opportunities for children's food education" (extracted factor) suggest that contact with children can play a significant role (Bonfert 2022; Morgan et al. 2018). In addition, "CSA farmers being present at farmers' markets" (extracted factor) influences consumer participation because consumers may perceive value beyond the availability of fresh produce through interactions with farmers at farmers' markets, which may lower their barriers to participating in CSAs (Farnsworth et al. 1996). This indicates that offering entry points through different approaches from CSA, such as programs in educational institutions or events at farmers' markets, can be effective.

For example, CSA could collaborate with school programs to educate children about sustainable agriculture and directly involve them in food production and environmental stewardship through experiential learning, thereby reaching more consumers. Previous study suggest that teaching kitchens, which partner with local farmers and use mobile kitchens to cook local ingredients while implementing a food education curriculum at schools, farmers' markets, or farms, promote interdisciplinary collaboration in healthcare, agriculture, and education and help build community ties (Cole, Pethan, and Evans 2023). Another study shows that CSAs are being used in anti-poverty initiatives, particularly in selling food to schools for students from low-income families, and some CSA farmers also sell as cooperatives within school districts (Flora and Bregendahl 2012). Such literature and cases suggest that integrating practical agricultural and nutrition education in collaboration with CSA farmers could allow CSA promoters to reach families and peers.

Our research shows that consumer decisions to participate in CSA are based on the premise that CSA is somewhat established in the area. Therefore, for widespread adoption of CSA and effective consumer outreach in specific areas, it is important to develop CSA promotion strategies that take into account the factors identified in our study. Promoters and farmers interested in practicing CSA should work to increase awareness of CSA and seek ways to communicate its value to consumers, with the goal of attracting a broader range of consumers. Collaborating with other institutions and related initiatives, making the value of CSA more tangible to consumers, and reducing barriers to participation, can encourage individual and family/peer decisions that lead to the successful practice of CSA.

#### 4.2.2. Suggestion for consumers

Our study found that in regions or countries with high levels of consumer concern about food, one factor contributing to participation is that CSAs "serve as one of the market channels" (extracted factor) and consumers "recognize CSA as a tool (a pure food acquisition route)" (extracted factor) similar to supermarkets and farmers' markets (Farnsworth et al. 1996; Feagan and Henderson 2009). In addition, many consumers emphasize fairness of share and price, and among the factors extracted, those related to food sourcing, such as quality, share content, and traceability, were the most numerous. From these results, it can be concluded that many CSA farms offer values that are attractive and understandable to consumers, namely the provision of fresh and safe food ingredients.

However, our paper highlights that the significance of CSA extends beyond being just a food supply source. It acts as a platform for fostering deeper connections between local producers and consumers, and for addressing environmental and social issues. This multifaceted role can be seen as the true essence of CSA. By participating in CSA, consumers build new values and trust through interactions with local producers and fellow consumers. This not only mitigates concerns about food and agriculture but also strengthens community bonds. Such connections can invigorate local communities, enhance satisfaction with participation, and create mutually beneficial scenarios for both producers and consumers. Therefore, consumers should understand the non-material values and multifaceted characteristics that participation in a CSA brings.

## 5. Conclusion

In our study, a scoping review was conducted to clarify the various factors involved in CSA participation and their relationships, and the consumer participation was theorized using the KJ method. Open coding was performed on 61 articles included in this review, and 306 factors were extracted. The relationships between these factors were then visualized and iteratively narrated to develop the theory. Through theory building, these factors were divided into the socio-cultural environment and the seesaw of gain and loss. The socio-cultural environment consists of four categories: individual, family/peer, local environment and community, and national agricultural frameworks, and the seesaw of gain and loss consists of two categories: gain and loss.

According to our theory, under the premise that consumers are embedded in a socio-cultural environment, their decisions to participate in CSA are influenced by the balance between expected gains and losses. This balance changes according to individuals' risk acceptance and their perception of gains and losses, leading to decision-making. This decision-making process is influenced by the embedded sociocultural environment, and it is suggested that discrepancies between expectations and post-participation experiences, and transformation of social capital in the CSA community, prompt updates in their decision-making process.

While the proposed theory offers a comprehensive analysis based on existing literature, including practitioner insights, also has limitations. Limitations of our study include that the scope of the scoping review is limited by the availability and accessibility of existing literature, and that the findings obtained may not be fully generalizable to all CSA contexts due to differences in cultural, economic, and geographic settings. In addition, the proposed theoretical framework needs to be validated and refined through further research, and the degree of influence of each participation factor needs to be compared, e.g., through questionnaire surveys, action research and case studies that examine the effectiveness and sustainability of CSA, as well as its impact on social capital. However, by thoroughly understanding the factors, backgrounds, and environments surrounding consumer participation, we can propose more effective initiatives and policies that can promote the spread and development of CSA.

In conclusion, by delineating the process of individual behavioral formation and decision-making, our findings pave the way for future research and practices aimed at fostering sustainable agriculture based on community engagement.

6. **CRediT authorship contribution statement**

**Sota Takagi:** Conceptualization, Methodology, Formal analysis, Investigation, Writing – Original Draft, Visualization, Supervision, Project administration. **Yusuke Numazawa**: Validation, Writing – Review & Editing, Funding acquisition. **Kentaro Katsube**: Validation, Writing – Review & Editing, Funding acquisition. **Wataru Omukai**: Validation, Writing – Review & Editing, Funding acquisition. **Miki Saijo:** Validation, Writing – Review & Editing. **Takumi Ohashi:** Methodology, Formal analysis, Validation, Writing – Visualization, Review & Editing. All authors contributed critically to the drafts and gave final approval for publication.

7. **Acknowledgments**

This study was conducted with the support of collaborative research funds from Eco-Pork Co., Ltd. The authors extend heartfelt gratitude to all participants who generously gave their time and insights for this research.

8. **Conflict of interest**

This research was conducted with research funding from Eco-Pork Co. Ltd., to which the second, third and fourth authors belong.

9. **Declaration of AI and AI-assisted technologies in the writing process**

In writing this paper, after preparing a full text draft in the lead author's non-English native language, the authors used OpenAI's artificial intelligence language model, ChatGPT, to prepare an English draft. After using this tool, the authors carefully reviewed and edited the generated content to ensure the flow, logic, and accuracy of the text, making additions as necessary. Therefore, full responsibility for the content of the publication rests with the authors.

**Appendix**

Some categories, concepts and sub concepts are included in more than one core category. <span style="color:red">Red</span> is the withdrawal factor, <span style="color:blue">blue</span> is the continuation factor, and black is the participation factor.

**Table A1 Socio-Cultural environment: Correspondence between all Extracted factors and Category, Concept, and Sub concept.**

| Core Category | Category | Concept | Sub concept | Extracted Factors | Ref. |
|---|---|---|---|---|---|
| (Individual) Knowledge and Experience | | Has experience and knowledge of CSA | having experience participating in CSA | Participation experience in CSA | [1] |
| | | | | Having been a member of CSA in the past | [2] |
| | | | | Having been a participant in CSA before | [3] |
| | | | Being aware of CSA | Being cognizant of CSA | [4] |
| | | | | Consumers personally knowing CSA farmers | [5] |
| | | Using Farmers' Markets | | Utilizing farmers' markets | [6] |
| | | | | High frequency of visits to farmers' markets | [1] |
| (Individual) Skills | Can/do cook fresh food at home | Ability to prepare food at home | Having adequate cooking skills and environment | Consumers possessing improvisational cooking skills. | [4] |
| | | | | <span style="color:blue">Having a good home cooking environment</span> | [7] |
| | | | Access to recipes | CSA farmers providing recipes for unfamiliar ingredients | [4] |
| | | | | Gaining knowledge about cooking and preservation methods for perishable items | [8] |
| | | | | Being able to get recipes | [9] |
| | | | | Enjoying cooking | [10] |
| | | | | <span style="color:red">Lacking knowledge about cooking</span> | [11] |
| | | Usually cook fresh food at home | Regularly purchasing and cooking fresh food | Vegetables being consumed within the household | [9] |
| | | | | Buying less processed food and more fresh food | [12] |
| | | | | High consumption of salads | [13] |
| | | | | Low consumption of processed food | [13] |
| | | | | High consumption of vegetables and fruits | [13] |
| | | | | Eating meals at home frequently | [2] |

| (Individual) Attitude towards Food | Interest in and commitment to food | Not compromise in food purchasing | Ability to avoid shopping at grocery stores | Eliminating the need to go to supermarkets | [14] |
|---|---|---|---|---|---|
| | | | | Having the intention to avoid shopping at grocery stores | [15] |
| | | | Ingredients being organically grown | Being organically cultivated | [16] |
| | | | | The ingredients being organic | [17] |
| | | | | Being organically cultivated | [3] |
| | | | | Preferring to buy organic vegetables | [18] |
| | | | | Preferring to eat organic vegetables | [19] |
| | | | | Being organic | [17] |
| | | | | Organically cultivated or using low-input farming methods. | [20] |
| | | | | Being free of pesticide | [21] |
| | | | | Organic vegetables being available | [22] |
| | | | | Organic vegetables being available | [23] |
| | | | | Being free of pesticide | [24] |
| | | | | Organic vegetables being available | [23] |
| | | | | Agricultural produce being organically grown / Livestock being fed organic feed | [25] |
| | | | | Purchasing organic vegetables | [26] |
| | | | | Purchasing organic vegetables | [10] |
| | | | | Wishing to increase the consumption of organic vegetables | [27] |
| | | | | Tendency to consume organic ingredients | [28] |
| | | | | The sales location being clean and well-organized | [17] |
| | | Interested in food | Wanting to enjoy food according to the season | Eating meals according to the season | [29] |
| | | | | Having the willingness to obtain seasonal vegetables | [6] |
| | | | | Being able to experience a sense of the season | [30] |

| | | | | Being able to obtain seasonal vegetables | [19] |
|---|---|---|---|---|---|
| | | | | Seasonality | [20] |
| | | | | High interest in food (valuing organic and local ingredients | [31] |
| | | | | Having a high interest in food | [32] |
| | | | | Being interested in special products, similar to those available in specialty stores | [31] |
| | | | | Believing that political, economic, and social factors are important in deciding where to purchase winter agricultural products | [26] |
| | | | | Having concerns about the quality of food ingredients | [33] |
| | | | | Having a tendency to engage in ethical consumption | [19] |
| | | | | CSA's agricultural products having a high market value, possibly due to being organic, fresh, etc. | [24] |
| | | Ability to Devote Effort to Food Preparation | Being able to spend time on food-related activities | Not being too busy | [2] |
| | | | | Having time for cooking | [11] |
| | | | | Not having time for cooking | [9] |
| | | | | Increasing time/effort spent on cooking vegetables at home | [34] |
| | | | | Saving time spent on grocery shopping | [29] |
| | | | | Spending money on grocery shopping | [2] |
| | | Ability to judge the reliability of ingredients: | The reliability (certification, qualitative) of ingredients being important | Existence of a quality certification system for ingredients | [35] |
| | | | | Trustworthiness of the ingredients being purchased | [5] |
| | | | Traceability of ingredients | Having reliable traceability | [36] |
| | | | | Valuing the traceability of ingredients | [29] |
| | | | | Consumers being interested in the cultivation methods and origin of ingredients | [6] |
| | | | | Traceability including the identity of the farmers | [37] |
| | | | | Knowing who grew the ingredients (seeing the face of the farmer) | [24] |

| | | | | Being able to view the process of obtaining the ingredients. | [38] |
|---|---|---|---|---|---|
| | | | | Having a traceability | [33] |
| | | | | Knowing the origin of food | [20] |
| | | | | Purchasing valuable, healthy, traceable food | [39] |
| (Individual) Attitude towards Environment | | Interest in environmental and social issues | Interest in sustainability | Interest in a sustainable society, including recycling and transportation methods | [18] |
| | | | | Interest in biodiversity | [18] |
| | | | | High frequency of recycling by consumers | [2] |
| | | | | Prioritizing support for sustainable agriculture | [29] |
| | | | | Interest in sustainability (environmental concerns, supporting farmers, preference for organic products) | [13] |
| | | | Practice of environmental activities | Recycling at home | [28] |
| | | | | Engaging in recycling and composting | [26] |
| | | | High environmental awareness | Being conscious of environmental considerations | [22] |
| | | | | Reducing carbon footprint | [20] |
| | | | | Ability to consider the environment | [23] |
| | | | | Vegetables not being packaged | [30] |
| | | | | Sustainability of ecology | [20] |
| | | | | Contributing to a better environment (environmental considerations) | [28] |
| | | | | Supporting environmental protection organizations | [28] |
| | | | | Willingness to support environmentally conscious practices | [27] |
| | | | | Tendency towards ethical consumption | [36] |
| | | | | Consumers being sensitive to environmental issues and local social problems | [40] |
| (Individual) Attitude towards Health | | Leading a life with health considerations | Expectations for health | Anticipating improvements in diet | [8] |
| | | | | Being able to perceive the value of health | [30] |

| | | | | Being a member of a gym | [2] |
|---|---|---|---|---|---|
| | | | | Having health consciousness (a factor that hasn't changed over time) | [22] |
| (Individual) Attitude towards Health / (Family/Peer) Family/Peer Environment and Value | | | High health consciousness | Consumers recognizing the legitimacy of a healthy lifestyle | [32] |
| | | | | Controlling salt intake | [17] |
| | | | | Contributing to one's own health or the health of family/peer members | [41] |
| (Family/Peer) Family/Peer Environment and Value | | Values towards food as a family/peer | Sharing values about CSA with a spouse | Receiving support from a spouse | [34] |
| | | | | Having common values about CSA with a spouse | [34] |
| | | | | Being in child-rearing | [41] |
| | | Leading a life with health considerations | | Having individuals with obesity or chronic diseases in the household | [17] |
| | | | Family/Peer composition | Not having children or teenagers at home | [26] |
| | | | | Having other adults in the household besides oneself | [10] |
| (Local Environment and Community) Connection and Attachment to the Community | | Word of mouth spreads easily | Knowing about CSA through word of mouth | Valuing word of mouth from newspapers or friends | [29] |
| | | | | Learning about CSA through word of mouth | [42] |
| | | | | Learning about CSA through word of mouth | [26] |
| | | | | Gaining and Valuing information about CSA through family and word of mouth. | [2] |
| | | | | CSA being introduced via word of mouth | [41] |
| | | | | The region being small and making it easy for reputation to spread | [39] |
| | | | | Knowing people who have tried F3B | [9] |
| | | Attachment to local farmers | Desire to consume local | Access to locally grown food | [43] |
| | | and region | ingredients | Seeking local produce | [33] |
| | | | | Ability to purchase local ingredients | [38] |
| | | | | Eating foods produced locally | [19] |
| | | | | Consuming local ingredients (due to environmental benefits) | [21] |
| | | | | Having the opportunity to buy locally produced and processed food | [39] |
| | | | | Being able to purchase local agricultural products | [23] |
| | | | | Access to local ingredients | [27] |
| | | | | Availability of a wider variety of fresh, local, organic foods | [27] |
| | | | | Being aware of the benefits of consuming local ingredients | [44] |
| | | | | Strong attachment to the community (e.g., local area and farmers) | [33] |
| (Local Environment and Community) Connection and Attachment to the Community/Culture | | | | Consumers having gratitude towards food and its place of origin | [25] |
| | | | | Farmers cultivating traditional ingredients or the region having such ingredients | [37] |
| (Local Environment and Community) Culture | | | Landscape preservation and environmental management | Conservation of open spaces | [20] |
| | | | | Management of the local environment | [20] |
| (Local Environment and Community) Local Norms | | Enthusiasm for supporting farmers and the local area | Consumers being enthusiastic about supporting the local area | CSA demonstrated to support the local economy | [5] |
| | | | | Enthusiastic about contributing to the local economy | [35] |
| | | | | Eager to support the development of farmers and the region | [35] |

| | | | | Supporting the local economy | [20] |
|---|---|---|---|---|---|
| | | | | Having the willingness to support the local economy | [45] |
| | | | | Being able to support the local community | [33] |
| | | | | Being able to support the local community | [38] |
| | | | | Willingness to support local businesses and value creation. | [27] |
| | | | Willingness to support producers | Willingness to support farmers in moving away from market principles | [46] |
| | | | | Ability to support alternative/organic agriculture | [41] |
| | | | Already having a relationship with farmers | CSA farmers being present at farmers' markets (consumers already knowing and purchasing from them) | [24] |
| | | | Concerns about the unstable financial state of farmers | Stabilization of CSA farmers' operations (support from consumers no longer being critical) | [47] |
| | | | Willingness to support local farmers | Being willing to support small-scale farmers | [23] |
| | | | | Ability to support farmers | [28] |
| | | | | Willingness to support the local food network | [48] |
| | | | | Desire to support local farmers | [6] |
| | | | | Supporting small-scale businesses | [36] |
| | | | | Supporting local farms (a factor that has strengthened over time) | [22] |
| (National agricultural framework conditions) Trends and Maturity of CSA | | | Availability of purchasing channels other than CSA | No chain stores of natural food stores | [27] |
| | | | | Existence of other options for purchasing ingredients (like supermarkets) | [42] |
| | | | Maturity as a channel | Recognize CSA as a tool (a pure food acquisition route) | [49] |
| | | | | Serve as one of the market channels | [24] |
| | CSA-related activities being active nationally or regionally. | | Connections of CSA with other institutions | CSA having collaborations with other institutions (universities, schools, social movements) | [50] |
| | | | | CSA having a network with other organizations | [51] |
| | | | Proximity to urban areas | Living in or near urban areas | [27] |

| | | | | Residing in urban areas | [23] |
|---|---|---|---|---|---|
| (National agricultural framework conditions) Support Systems and Organizations | | | Support from local governments and related organizations | Funding agencies having interest in the region | [3] |
| | | | | High agricultural demand in that country or region | [52] |
| | | | | Local governments attempting to increase local food consumption. | [3] |
| (National agricultural framework conditions) Agriculture Policy | | CSA already being practiced in the country or region. | High awareness of CSA by the country or local government | Support and approval for CSA by the country or region | [53] |
| | | | | Well-established national policies related to CSA. | [54] |
| | | | | Multiple CSAs existing in the region. | [28] |
| | | | | Several CSAs being implemented in the region. | [3] |
| | | | | No competition among CSAs in the region. | [16] |

**Table A2 Seesaw of Gain and Loss: Correspondence between all Extracted factors and Category, Concept, and Sub concept.**

| Core Category | Category | Concept | Sub concept | Extracted Factors | Ref. |
|---|---|---|---|---|---|
| (Gain) Food Education and Learning Opportunities | | Expectation of opportunities for food education and learning | Opportunities to learn about food | Places to learn about food | [52] |
| | | | | Ability to acquire knowledge | [36] |
| | | | | Ability to gain knowledge about food and awareness of food handling | [25] |
| | | | | Learning opportunities about food production | [17] |
| | | | Opportunities for food education | Prospect of improved education | [26] |
| | | | | Learning orientation | [49] |
| | | | Learning opportunities in agriculture | Desire to learn about agriculture and ecology (higher in households with children) | [27] |
| | | | | Ability to learn about agriculture | [8] |
| | | | | Desire to acquire knowledge about local soil resource management and local knowledge | [27] |
| | | | | Learning about seasonality, more sustainable diets, trying new things with seasonal produce, and using new vegetables | [55] |
| | | | | Having opportunities to learn | [55] |
| | | | | Opportunities for children's food education | [5] |
| (Gain) Contribution to Environmental and Social Issues | | Expectation of solving environmental and social issues through CSA participation | Recognition of CSA's impact on the environment | CSA's sustainability in environmental aspects | [52] |
| | | | | Perception that CSA has a lower environmental impact | [2] |
| | | | Critical of the conventional food production and distribution system | Critical of conventional agriculture | [46] |
| | | | | Critical of the current food production system | [46] |
| | | | | Critical of the traditional economic system | [36] |
| | | | | Absence of intensive cultivation practices | [30] |
| | | | | Willingness to support a food system without merchandisers | [24] |
| | | | | Negative view of the traditional production system | [23] |
| | | | | Rebuilding the direct channel between urban and rural areas | [24] |

| (Gain) Connections with People and Nature | | | | | |
|---|---|---|---|---|---|
| | | | Feeling a connection with non-human elements | Connection with places, local ecosystems | [20] |
| | | | | Encounters with non-human life (physical/virtual) | [37] |
| | | | Opportunities to share experiences | Talking about CSA farms and learned knowledge with others | [55] |
| | | | | Communication among participants | [56] |
| | Expectation of human connections through CSA participation | Ability to connect with farmers | Having a shared consciousness with farmers | Established trust between consumers and producers | [57] |
| | | | | Shared values with farmers | [53] |
| | | | Direct communication with farmers | Ability to communicate directly with farmers | [35] |
| | | | | Difficulty in communicating with CSA staff and farmers | [11] |
| | | | | Not sharing the values of CSA | [30] |
| | | | | Personal connection with farmers | [20] |
| | | Expectation of interaction opportunities through CSA participation | Opportunities to deepen relationships with close ones through CSA | Opportunities to deepen relationships with close ones through CSA (such as family and friends) | [56] |
| | | | | Opportunities to deepen specific relationships and to communicate with people | [56] |
| | | | Opportunities for social involvement through CSA | Opportunities for face-to-face communication | [56] |
| | | | | Possibility of social interaction | [36] |
| | | | Expectation of community formation | Meeting like-minded people | [43] |
| | | | | Ability to build a sense of community | [19] |
| | | | | Formation of a community among members (higher among consumers who have participated in farm activities) | [27] |
| | | | | Seeking a stronger sense of community (importance diminishes over time) | [22] |
| | | | | Creation/maintenance of community | [20] |
| | | | | Ability to form strong communities | [43] |
| | | | | Existence of a formed community | [55] |
| | | | | High quality and satisfaction of the community | [33] |
| | Variety in the operational | | | Absence of animal husbandry at the farm | [50] |

| (Gain) Variety of Ingredients | styles of CSA farms that can be participated in | | | The introduction of specific options that one desires (such as trying new vegetables that have never been tried before, healthier choices than going to the store, supporting farmers) | [55] |
|---|---|---|---|---|---|
| | Expectation of a variety of high-quality food ingredients | A wide range of share options available | The importance of the content of the available share | Ability to obtain fruits | [4] |
| | | | | Option to purchase meat as well | [37] |
| | | | | Inclusion of fruits and flowers | [16] |
| | | | | Inclusion of meat in addition to vegetables in the share | [3] |
| | | | Access to unfamiliar food ingredients | Unfamiliarity with the food ingredients | [23] |
| | | | | Ability to obtain uncommon food ingredients | [37] |
| | | | | Access to unfamiliar food ingredients | [48] |
| | | | | Ability to learn about new food ingredients | [8] |
| | | | Importance of the quantity of the share | Adequate quantity of produce | [54] |
| | | | Importance of the range of share options | Lack of appropriate diversity in products | [11] |
| | | | | Wide range of product options (such as half-share/full-share) | [5] |
| | | | | Abundance of vegetable options | [40] |
| | | | | Rich variety and quantity of vegetables | [55] |
| | | | | Ability to choose food ingredients based on the above | [4] |
| | | | | Inability to choose vegetables in the share | [11] |
| | | | Importance of the frequency of the share | Many (appropriate) months of availability per year (number of share times) | [16] |
| | | | | Contract duration ranging from 4 weeks to 12 weeks | [1] |
| | | | Variety in the share | Variety in food ingredients | [45] |
| | | | | Safety and diversity of food types | [28] |
| | | | | Flexibility in the choice of vegetables (in terms of quantity and type) | [38] |
| | | | Ease of freely choosing | Ability to sample recipes using products from that week's box | [9] |

| | | | | Ability to compare and choose food ingredients oneself (in the context of China) | [35] |
|---|---|---|---|---|---|
| | | | share ingredients | Familiarity with the vegetables (especially interest in summer vegetables) | [4] |
| | | | | Inclusion of lightly processed vegetables | [9] |
| | | | | Food ingredients being ordinary | [26] |
| | | The importance of high-quality and fresh ingredients | High-quality food ingredients | Lack of access to quality food ingredients in the surrounding area | [58] |
| | | | | Availability of high-quality and fresh food products | [41] |
| | | | | High quality of the products | [11] |
| | | | | Ability to obtain high-quality food ingredients | [48] |
| | | | | High quality of food ingredients | [28] |
| | | | | Vegetables fresher than those available at supermarkets | [9] |
| | | | | Good quality of food ingredients | [36] |
| | | | | Preference for high-quality food ingredients | [7] |
| | | | | High quality of food ingredients | [7] |
| | | | | High-quality products | [45] |
| | | | Desire for fresh ingredients | Consumer desire for fresh food ingredients | [15] |
| | | | | Interest in fresh food ingredients | [31] |
| | | | | Availability of fresh vegetables | [22] |
| | | | | Ability to eat freshly harvested vegetables and fruits | [19] |
| | | | | Access to fresh and healthy food | [28] |
| | | | | Availability of fresh vegetables and fruits | [43] |
| | | | | Freshness, taste, and nutritional content | [20] |
| | | | | Access to fresh and nutritious agricultural produce | [23] |
| | | | | Ability to obtain fresh and delicious food | [25] |
| | | | | Aesthetically pleasing appearance of vegetables | [40] |
| (Loss) Complicated Relationships | Variety in the operational styles of CSA farms that can be participated in | Choices in forms of community participation | Many restrictions within the community | Overly strong relationships within the community | [47] |
| | | | | Strict constraints within the community | [59] |
| | | | | A lot of work on the farm | [47] |
| | | | | Availability of flexible employment opportunities | [42] |
| | | | | Choices in types of participation in CSA | [60] |
| (Loss) Money | | | Consumers not incurring losses | Mechanisms to mitigate risks in case of crop failure (such as refunds) | [3] |
| | | | | Alternatives in case of crop failure | [54] |
| | | Financial and time cost-performance | Feeling the fairness in the content and price of the share | Ability to obtain various types of vegetables at a uniform price | [9] |
| | | | | Large share sizes with affordable prices | [61] |
| | | | | Prices being within 20% higher than supermarkets | [61] |
| | | | | Being able to know the breakdown of the food ingredients provided before making the payment | [14] |
| | | | | Feeling that the price of the share is reasonable | [38] |
| | | | | Appropriate pricing of vegetables | [22] |
| | | | | Membership fees not being high | [26] |
| | | | | Appropriateness of the share price | [44] |
| | | | | Reduction in food expenses | [8] |
| (Loss) Time | | | Geographical location of the pickup point | No physical effort required to access the food ingredients | [38] |
| | | | | Pickup points not located in inconvenient places | [9] |
| | | | | Pickup points located along the consumer's travel route | [58] |
| | | | | Collection points or farms located along the typical travel route | [5] |
| | | | | Farms located in areas with high land value | [54] |
| | | | | Pickup points in inconvenient locations | [11] |

| | | | High accessibility | Implementation of online marketing | [39] |
| --- | --- | --- | --- | --- | --- |
| | | | | High accessibility (from the perspective of time and location) | [44] |

**Table A3 Others: Correspondence between all Extracted factors and Category, Concept, and Sub concept.**

| Core Category | Category | Concept | Sub concept | Extracted Factors | Ref. |
|---|---|---|---|---|---|
| Risk Acceptance | | | Recognition of the benefits of taking risks | Ability to share financial risks | [43] |
| | | | | Willingness to share risks | [1] |
| Participation | | Consumer involvement in community management | Consumers' ability to be involved in the operation of CSA farms | Ability to be involved beyond being just a consumer | [47] |
| | | | | Having the authority to make decisions on CSA farm operations | [52] |
| | | | | Safe environment for farm volunteer work | [60] |
| | | | | Opportunities for volunteer activities | [43] |
| | | | Desire to participate in production activities | Desire to be involved in the cultivation of their own food | [27] |
| | | | | Ability to participate in farm events | [43] |
| | | | Consumers' participation in collective activities | Consumers being active in group activities | [40] |
| | | | | Active participation in political activities | [46] |
| | | | | Having a functional cooperative orientation | [49] |
| Demographic and Psychographic | | | | Engaging in home gardening | [6] |
| | | | | Gardening activities | [10] |
| | | | | Having someone else choose vegetables for them | [24] |
| | | | Stability of consumers' jobs | Job stability | [17] |
| | | | | Established career | [41] |
| | | | | Low likelihood of moving | [41] |
| | | | High educational attainment of consumers | Higher education | [22] |
| | | | | Higher education | [23] |
| | | | | Higher education | [28] |
| | | | | Higher academic qualifications | [21] |
| | | | | Receiving higher education | [17] |

| | | | | Higher education | [31] |
|---|---|---|---|---|---|
| | | | High-income status | Belonging to the middle class | [23] |
| | | | | Relatively high income | [54] |
| | | | | High income | [19] |
| | | | | Relatively high income | [13] |
| | | | | High household income | [22] |
| | | | | High household earnings | [21] |
| | | | | Young consumers with high income | [40] |
| | | | | Being a high-income earner | [29] |
| | | | Easier participation for women | Being a woman | [9] |
| | | | | Consumers being female | [31] |
| | | | Relatively young age group | Being middle-aged (30–59 years) | [31] |
| | | | | Younger generation | [29] |
| | | | | Being Caucasian | [9] |
| | | | | Being Caucasian | [23] |